\documentclass[twocolumn,traditabstract]{aa}

\usepackage[latin9]{inputenc}
\usepackage{amsmath}
\usepackage{amssymb}
\usepackage{natbib}
\usepackage{graphicx}
\bibpunct{(}{)}{;}{a}{}{,} 
\newcommand{\cG}{{\cal G}}

\newcommand{\br}{{\mathbf{r}}}
\newcommand{\bi}{{\mathbf{i}}}
\newcommand{\bj}{{\mathbf{j}}}
\newcommand{\bk}{{\mathbf{k}}}
\newcommand{\uu}{{\mathbf{u}}}

\newcommand{\dpp} {\!\cdot\!}
\newcommand{\abs}[1]{{\left|{#1}\right|}}

\newcommand{\al} {\alpha}
\renewcommand{\ga} {\gamma}
\newcommand{\om} {\omega}
\newcommand{\Om} {\Omega}
\newcommand{\cF} {{\cal F}}
\newcommand{\e}{{\mathbf{e}}}

\def\norm#1{\left\Vert#1\right\Vert}

\renewcommand{\i} {{\mathbf{i}}}
\newcommand{\cA}{{\cal A}}

\def\crm{\cr\noalign{\medskip}}

\def\m@th{\mathsurround=0pt}
\def\EQM#1{\vcenter{\normalbaselines\m@th
    \ialign{${\displaystyle ##}$\hfil&&\ ${\displaystyle ##}$\hfil\crcr
    \mathstrut\crcr\noalign{\kern-\baselineskip}
    \noalign{\smallskip}
    #1\crcr\mathstrut\crcr\noalign{\kern-\baselineskip}}}}
\newcommand{\Frac}[2]{{{\displaystyle\strut#1}\over{\displaystyle\strut#2}}}
\newcommand{\be}{\begin{equation}}
\newcommand{\ee}{\end{equation}}

\def\Dron#1#2{\frac{\partial#1}{\partial#2}}

\newcommand{\bpm}{\begin{pmatrix}}
\newcommand{\epm}{\end{pmatrix}}

\newcommand{\cC} {{\cal C}}
\newcommand{\cQ} {{\cal Q}}
\newcommand{\cZ} {{\cal Z}}

\newcommand{\nul}[1]{}
\newcommand{\tmu} {\mu_*}
\newcommand{\tnu} {\nu_*}
\newcommand{\mybf}[1]{{#1}}

\newcommand{\tabii}{
\begin{table*}
\begin{tabular}{r l}
\hline\hline
$F_{0}$ &= $1$ \\[10pt]
$F_{1}$ &= $   - b' \sin(u+u')
 +b \cos(u+u')
 - a' \sin(u-u')
 + a \cos(u-u') $\\[10pt]
$F_{2}$ &= $ 
 -\Frac{1}{2}
 +\Frac{3}{4} b'^2
 +\Frac{3}{4} b^2
 +\Frac{3}{4} a'^2
 +\Frac{3}{4} a^2 $\\
 &$+(\Frac{3}{2} b a
 -\Frac{3}{2} b' a') \cos(2 u)
 -(\Frac{3}{2} b a'  
 +\Frac{3}{2} b' a )\sin(2 u)
  +(\Frac{3}{2} b' a' 
 +\Frac{3}{2} b a) \cos(2 u')
 +(\Frac{3}{2} b a' 
 -\Frac{3}{2} b' a )\sin(2 u')$ \\
&$ +(\Frac{3}{4} b^2 
 -\Frac{3}{4} b'^2) \cos(2 u+2 u')
 +(\Frac{3}{4} a^2 
 -\Frac{3}{4} a'^2) \cos(2 u-2 u')
 -\Frac{3}{2} b' b \sin(2 u+2 u')
 +\Frac{3}{2} a' a \sin(-2 u+2 u')$\\[10pt]
 $F_{3}$ &= $ 
 +(\Frac{15}{4} b'^2 a  
 +\Frac{15}{4} b^2 a   
 +\Frac{15}{8} a'^2 a 
 +\Frac{15}{8} a^3    
 -\Frac{3}{2} a      ) \cos(-u+u')
 +(\Frac{15}{4} b a'^2  
 +\Frac{15}{4} b a^2   
 -\Frac{3}{2} b        
 +\Frac{15}{8} b'^2 b  
 +\Frac{15}{8} b^3    )\cos(u+u')$\\
 &$+(\Frac{3}{2} b'   
 -\Frac{15}{8} b'^3   
 -\Frac{15}{8} b' b^2 
 -\Frac{15}{4} b' a'^2
 -\Frac{15}{4} b' a^2  )\sin(u+u')
 +(\Frac{15}{8} a' a^2 
 -\Frac{3}{2} a'       
 +\Frac{15}{4} b'^2 a' 
 +\Frac{15}{4} b^2 a'  
 +\Frac{15}{8} a'^3  )  \sin(-u+u')$\\
&$ 
 +(\Frac{15}{4} b' b a'  
 -\Frac{15}{8} b'^2 a  
 +\Frac{15}{8} b^2 a )\cos(u+3 u')
 +(\Frac{15}{8} b^2 a'  
 -\Frac{15}{8} b'^2 a'  
 -\Frac{15}{4} b' b a ) \sin(u+3 u')$\\
  &$+(\Frac{15}{8} b a^2  
 -\Frac{15}{8} b a'^2  
 +\Frac{15}{4} b' a' a) \cos(-u+3 u')
 +(\Frac{15}{8} b' a'^2 
 +\Frac{15}{4} b a' a   
 -\Frac{15}{8} b' a^2 ) \sin(-u+3 u')$\\
&$ 
 +(\Frac{15}{8} b^2 a  
 -\Frac{15}{4} b' b a'  
 -\Frac{15}{8} b'^2 a )\cos(3 u+u')
 +(\Frac{15}{8} b'^2 a'  
 -\Frac{15}{8} b^2 a'  
 -\Frac{15}{4} b' b a )\sin(3 u+u')$\\
 &$+(\Frac{15}{8} b a^2  
 -\Frac{15}{8} b a'^2  
 -\Frac{15}{4} b' a' a )\cos(-3 u+u')
 +(\Frac{15}{4} b a' a  
 +\Frac{15}{8} b' a^2  
 -\Frac{15}{8} b' a'^2 )\sin(-3 u+u')$\\
&$ 
 +(\Frac{5}{8} b^3  
 -\Frac{15}{8} b'^2 b) \cos(3 u+3 u')
 +(\Frac{5}{8} b'^3  
 -\Frac{15}{8} b' b^2 )\sin(3 u+3 u')$\\
&$ +(\Frac{5}{8} a^3 
 -\Frac{15}{8} a'^2 a )\cos(-3 u+3 u')
 +(\Frac{15}{8} a' a^2
 -\Frac{5}{8} a'^3 )\sin(-3 u+3 u')$\\
\hline
\end{tabular} 
\caption{\label{eq.fnmunu}Tisserand functions for the spatial case in a fixed reference frame. We have $u=v+\om,u'=v'+\om'$ while  $a,a',b,b'$ 
depend on the nodes and inclinations and  are given in (\ref{eq.aabb}).}
\end{table*}
}

\newcommand{\tabb}{
\begin{table*}
\begin{tabular}{r l}
\hline\hline
$F_{0}$ &= $1$ \\[10pt]
$F_{1}$ &= $   \mu \cos(u-u') + \nu \cos(u+u')$\\[10pt]
$F_{2}$ &= $ 
 - \Frac{1}{2}
 + \Frac{3}{4} \nu^2
 + \Frac{3}{4} \mu^2
 + \Frac{3}{4} \mu^2 \cos(2u-2u')
 + \Frac{3}{4} \nu^2 \cos(2u+2u')
 + \Frac{3}{2} \nu \mu \cos(2u)
 + \Frac{3}{2} \nu \mu \cos(2u')$\\[15pt]
$F_{3}$ &=$
 \left(- \Frac{3}{2} \mu  + \Frac{15}{4} \nu^2 \mu   + \Frac{15}{8} \mu^3 \right)  \cos(u-u')
 +\left(- \Frac{3}{2} \nu    + \Frac{15}{8} \nu^3   + \Frac{15}{4} \nu \mu^2\right) \cos(u+u')$\\
 &$+ \Frac{15}{8} \nu^2 \mu \cos(3u+u')
 + \Frac{15}{8} \nu^2 \mu \cos(u+3u')$\\
 &$+ \Frac{15}{8} \nu \mu^2 \cos(3u-u')
 + \Frac{15}{8} \nu \mu^2 \cos(u-3u')
 + \Frac{5}{8} \mu^3 \cos(3u-3u')
 + \Frac{5}{8} \nu^3 \cos(3u+3u')$\\[15pt]
$F_{4}$ &=$
 + \Frac{3}{8}
 - \Frac{15}{8} \nu^2
 + \Frac{105}{64} \nu^4
 - \Frac{15}{8} \mu^2
 + \Frac{105}{16} \nu^2 \mu^2
 + \Frac{105}{64} \mu^4$\\
&$ +\left(
 - \Frac{15}{8} \mu^2  
 + \Frac{105}{16} \nu^2 \mu^2  
 + \Frac{35}{16} \mu^4\right) \cos(2u-2u')
 +\left(
 - \Frac{15}{8} \nu^2  
 + \Frac{35}{16} \nu^4  
 + \Frac{105}{16} \nu^2 \mu^2 \right)\cos(2u+2u')$\\
 &$+\left(
 - \Frac{15}{4} \nu \mu 
 + \Frac{105}{16} \nu^3 \mu  
 + \Frac{105}{16} \nu \mu^3\right) \cos(2u)
 +\left(
 - \Frac{15}{4} \nu \mu  
 + \Frac{105}{16} \nu^3 \mu  
 + \Frac{105}{16} \nu \mu^3\right) \cos(2u')$\\
 &$+ \Frac{35}{64} \mu^4 \cos(4u-4u')
 + \Frac{35}{64} \nu^4 \cos(4u+4u')
 + \Frac{35}{16} \nu \mu^3 \cos(4u-2u')
 + \Frac{35}{16} \nu \mu^3 \cos(2u-4u')$\\
 &$+ \Frac{105}{32} \nu^2 \mu^2 \cos(4u)
 + \Frac{105}{32} \nu^2 \mu^2 \cos(4u')
 + \Frac{35}{16} \nu^3 \mu \cos(4u+2u')
 + \Frac{35}{16} \nu^3 \mu \cos(2u+4u')$\\[15pt]
$F_{5}$ &=$ 
  \left(
 +\Frac{15}{8} \mu  
 -\Frac{105}{8} \nu^2 \mu  
 +\Frac{945}{64} \nu^4 \mu  
 -\Frac{105}{16} \mu^3  
 +\Frac{945}{32} \nu^2 \mu^3  
 +\Frac{315}{64} \mu^5\right)  \cos(u-u')$\\
 &$+\left(
 +\Frac{15}{8} \nu  
 -\Frac{105}{16} \nu^3  
 +\Frac{315}{64} \nu^5 
 -\Frac{105}{8} \nu \mu^2  
 +\Frac{945}{32} \nu^3 \mu^2  
 +\Frac{945}{64} \nu \mu^4 \right) \cos(u+u')$\\
 &$+\left(
 -\Frac{105}{16} \nu^2 \mu  
 +\Frac{315}{32} \nu^4 \mu  
 +\Frac{945}{64} \nu^2 \mu^3\right) \cos(u+3u') 
  +\left(
 -\Frac{105}{16} \nu \mu^2  
 +\Frac{945}{64} \nu^3 \mu^2  
 +\Frac{315}{32} \nu \mu^4 \right)\cos(u-3u')$\\
 &$+\left(
 -\Frac{105}{16} \nu^2 \mu  
 +\Frac{315}{32} \nu^4 \mu  
 +\Frac{945}{64} \nu^2 \mu^3 \right) \cos(3u+u') 
  +\left(
 -\Frac{105}{16} \nu \mu^2  
 +\Frac{945}{64} \nu^3 \mu^2  
 +\Frac{315}{32} \nu \mu^4  \right)\cos(3u-u')$\\
 &$+\left(
 -\Frac{35}{16} \nu^3  
 +\Frac{315}{128} \nu^5  
 +\Frac{315}{32} \nu^3 \mu^2 \right)\cos(3u+3u') 
  +\left(
 -\Frac{35}{16} \mu^3  
 +\Frac{315}{32} \nu^2 \mu^3 
 +\Frac{315}{128} \mu^5 \right)\cos(3u-3u')$\\
 &$+\Frac{315}{64} \nu^3 \mu^2 \cos(u+5u')
 +\Frac{315}{64} \nu^2 \mu^3 \cos(u-5u')
 +\Frac{315}{64} \nu^3 \mu^2 \cos(5u+u')
 +\Frac{315}{64} \nu^2 \mu^3 \cos(5u-u')
 +\Frac{315}{128} \nu^4 \mu \cos(3u+5u')$\\
 &$+\Frac{315}{128} \nu \mu^4 \cos(3u-5u')
 +\Frac{315}{128} \nu^4 \mu \cos(5u+3u')
 +\Frac{315}{128} \nu \mu^4 \cos(5u-3u')
 +\Frac{63}{128} \nu^5 \cos(5u+5u')
 +\Frac{63}{128} \mu^5 \cos(5u-5u')$\\
\hline
\end{tabular} 
\caption{\label{eq.fnmunuii}Tisserand functions for the spatial case. We have $u=v+\om,u'=v'+\om'$ while  $\mu=\cos^2 (J/2)$, 
$\nu=\sin^2 (J/2)$, where $J$ is the mutual inclination.}
\end{table*}
}

\newcommand{\SSa}{ \vphantom{\rule[-7pt]{0pt}{22pt}}}
\newcommand{\SSb}{ \vphantom{\rule[0pt]{0pt}{10pt}}}
\newcommand{\SSc}{ \vphantom{\rule[-5pt]{0pt}{15pt}}}

\newcommand{\tabc}{
\begin{table}[h!]
\begin{tabular}{|cc| l | cc| l|}
\hline
\SSa $n$ &$m$ &\hfil $X_0^{n,m}$ & $n$ &$m$ &\hfil $X_0^{n,m}$\\
\hline
\SSb$0$ & $0$ & $1$ &                                                           $7$ & $0$ & $ +1+14e^{2}+{105}/{4}\,e^{4}+{35}/{4}\,e^{6}+{35}/{128}\,e^{8}$  \\[1pt]
\SSb$1$ & $0$ & $ +1+{1}/{2}\,e^{2}$ &                                           $7$ & $1$ & $ -{9}/{2}\,e-{189}/{8}\,e^{3}-{315}/{16}\,e^{5}-{315}/{128}\,e^{7}$  \\
\SSb$1$ & $1$ & $ -{3}/{2}\,e$ &                                                 $7$ & $2$ & $ +{45}/{4}\,e^{2}+{225}/{8}\,e^{4}+{675}/{64}\,e^{6}+{45}/{128}\,e^{8}$  \\
\SSb$2$ & $0$ & $ +1+{3}/{2}\,e^{2}$ &                                                          $7$ & $3$ & $ -{165}/{8}\,e^{3}-{825}/{32}\,e^{5}-{495}/{128}\,e^{7}$  \\
\SSb$2$ & $1$ & $ -2e-{1}/{2}\,e^{3}$ &                                                         $7$ & $4$ & $ +{495}/{16}\,e^{4}+{297}/{16}\,e^{6}+{99}/{128}\,e^{8}$  \\
\SSb$2$ & $2$ & $ +{5}/{2}\,e^{2}$ &                                                            $7$ & $5$ & $ -{1287}/{32}\,e^{5}-{1287}/{128}\,e^{7}$  \\
\SSb$3$ & $0$ & $ +1+3e^{2}+{3}/{8}\,e^{4}$ &                                                   $7$ & $6$ & $ +{3003}/{64}\,e^{6}+{429}/{128}\,e^{8}$  \\
\SSb$3$ & $1$ & $ -{5}/{2}\,e-{15}/{8}\,e^{3}$ &                                                $7$ & $7$ & $ -{6435}/{128}\,e^{7}$ \\
\SSb$3$ & $2$ & $ +{15}/{4}\,e^{2}+{5}/{8}\,e^{4}$ &                                                $8$ & $0$ & $ +1+18e^{2}+{189}/{4}\,e^{4}+{105}/{4}\,e^{6}+{315}/{128}\,e^{8}$\\ 
\SSb$3$ & $3$ & $ -{35}/{8}\,e^{3}$ &                                                             $8$ & $1$ & $ -5e-35e^{3}-{175}/{4}\,e^{5}-{175}/{16}\,e^{7}-{35}/{128}\,e^{9}$\\ 
\SSb$4$ & $0$ & $ +1+5e^{2}+{15}/{8}\,e^{4}$ &                                                    $8$ & $2$ & $ +{55}/{4}\,e^{2}+{385}/{8}\,e^{4}+{1925}/{64}\,e^{6}+{385}/{128}\,e^{8}$\\ 
\SSb$4$ & $1$ & $ -3e-{9}/{2}\,e^{3}-{3}/{8}\,e^{5}$ &                                              $8$ & $3$ & $ -{55}/{2}\,e^{3}-{825}/{16}\,e^{5}-{495}/{32}\,e^{7}-{55}/{128}\,e^{9}$\\ 
\SSb$4$ & $2$ & $ +{21}/{4}\,e^{2}+{21}/{8}\,e^{4}$ &                                             $8$ & $4$ & $ +{715}/{16}\,e^{4}+{715}/{16}\,e^{6}+{715}/{128}\,e^{8}$\\ 
\SSb$4$ & $3$ & $ -7e^{3}-{7}/{8}\,e^{5}$ &                                                       $8$ & $5$ & $ -{1001}/{16}\,e^{5}-{1001}/{32}\,e^{7}-{143}/{128}\,e^{9}$\\ 
\SSb$4$ & $4$ & $ +{63}/{8}\,e^{4}$ &                                                             $8$ & $6$ & $ +{5005}/{64}\,e^{6}+{2145}/{128}\,e^{8}$\\ 
\SSb$5$ & $0$ & $ +1+{15}/{2}\,e^{2}+{45}/{8}\,e^{4}+{5}/{16}\,e^{6}$ &                           $8$ & $7$ & $ -{715}/{8}\,e^{7}-{715}/{128}\,e^{9}$\\ 
\SSb$5$ & $1$ & $ -{7}/{2}\,e-{35}/{4}\,e^{3}-{35}/{16}\,e^{5}$ &                                 $8$ & $8$ & $ +{12155}/{128}\,e^{8}$\\ 
\SSb$5$ & $2$ & $ +7e^{2}+7e^{4}+{7}/{16}\,e^{6}$ &                                               $9$ & $0$ & $ +1+{45}/{2}\,e^{2}+{315}/{4}\,e^{4}+{525}/{8}\,e^{6}+{1575}/{128}\,e^{8}+{63}/{256}\,e^{10}$\\ 
\SSb$5$ & $3$ & $ -{21}/{2}\,e^{3}-{63}/{16}\,e^{5}$ &                                          $9$ & $1$ & $ -{11}/{2}\,e-{99}/{2}\,e^{3}-{693}/{8}\,e^{5}-{1155}/{32}\,e^{7}-{693}/{256}\,e^{9}$\\ 
\SSb$5$ & $4$ & $ +{105}/{8}\,e^{4}+{21}/{16}\,e^{6}$ &                                         $9$ & $2$ & $ +{33}/{2}\,e^{2}+77e^{4}+{1155}/{16}\,e^{6}+{231}/{16}\,e^{8}+{77}/{256}\,e^{10}$\\ 
\SSb$5$ & $5$ & $ -{231}/{16}\,e^{5}$ &                                                           $9$ & $3$ & $ -{143}/{4}\,e^{3}-{3003}/{32}\,e^{5}-{3003}/{64}\,e^{7}-{1001}/{256}\,e^{9}$\\ 
\SSb$6$ & $0$ & $ +1+{21}/{2}\,e^{2}+{105}/{8}\,e^{4}+{35}/{16}\,e^{6}$ &                         $9$ & $4$ & $ +{1001}/{16}\,e^{4}+{3003}/{32}\,e^{6}+{3003}/{128}\,e^{8}+{143}/{256}\,e^{10}$\\ 
\SSb$6$ & $1$ & $ -4e-15e^{3}-{15}/{2}\,e^{5}-{5}/{16}\,e^{7}$ &                                $9$ & $5$ & $ -{3003}/{32}\,e^{5}-{5005}/{64}\,e^{7}-{2145}/{256}\,e^{9}$\\ 
\SSb$6$ & $2$ & $ +9e^{2}+15e^{4}+{45}/{16}\,e^{6}$ &                                           $9$ & $6$ & $ +{1001}/{8}\,e^{6}+{429}/{8}\,e^{8}+{429}/{256}\,e^{10}$\\ 
\SSb$6$ & $3$ & $ -15e^{3}-{45}/{4}\,e^{5}-{9}/{16}\,e^{7}$ &                                     $9$ & $7$ & $ -{2431}/{16}\,e^{7}-{7293}/{256}\,e^{9}$\\ 
\SSb$6$ & $4$ & $ +{165}/{8}\,e^{4}+{99}/{16}\,e^{6}$ &                                           $9$ & $8$ & $ +{21879}/{128}\,e^{8}+{2431}/{256}\,e^{10}$\\ 
\SSb$6$ & $5$ & $ -{99}/{4}\,e^{5}-{33}/{16}\,e^{7}$ &                                        $9$ & $9$ & $ -{46189}/{256}\,e^{9}$\\ 
\SSb$6$ & $6$ & $ +{429}/{16}\,e^{6}$ &                                                       $10$ & $0$ & $ +1+{55}/{2}\,e^{2}+{495}/{4}\,e^{4}+{1155}/{8}\,e^{6}+{5775}/{128}\,e^{8}+{693}/{256}\,e^{10}$\\ 
\SSb&&&                     $10$ & $1$ & $ -6e-{135}/{2}\,e^{3}-{315}/{2}\,e^{5}-{1575}/{16}\,e^{7}-{945}/{64}\,e^{9}-{63}/{256}\,e^{11}$\\ 
\SSb&&&                  $10$ & $2$ & $ +{39}/{2}\,e^{2}+117e^{4}+{2457}/{16}\,e^{6}+{819}/{16}\,e^{8}+{819}/{256}\,e^{10}$\\ 
\SSb&&&                   $10$ & $3$ & $ -{91}/{2}\,e^{3}-{637}/{4}\,e^{5}-{1911}/{16}\,e^{7}-{637}/{32}\,e^{9}-{91}/{256}\,e^{11}$\\ 
\SSb&&&                     $10$ & $4$ & $ +{1365}/{16}\,e^{4}+{5733}/{32}\,e^{6}+{9555}/{128}\,e^{8}+{1365}/{256}\,e^{10}$\\ 
\SSb&&&                  $10$ & $5$ & $ -{273}/{2}\,e^{5}-{1365}/{8}\,e^{7}-{585}/{16}\,e^{9}-{195}/{256}\,e^{11}$\\ 
\SSb&&&                    $10$ & $6$ & $ +{1547}/{8}\,e^{6}+{1105}/{8}\,e^{8}+{3315}/{256}\,e^{10}$\\ 
\SSb&&&                      $10$ & $7$ & $ -{1989}/{8}\,e^{7}-{5967}/{64}\,e^{9}-{663}/{256}\,e^{11}$\\ 
\SSb&&&                    $10$ & $8$ & $ +{37791}/{128}\,e^{8}+{12597}/{256}\,e^{10}$\\ 
\SSb&&&                    $10$ & $9$ & $ -{20995}/{64}\,e^{9}-{4199}/{256}\,e^{11}$\\ 
\SSc&&&                    $10$ & $10$ & $ +{88179}/{256}\,e^{10}$\\ 
\hline
\end{tabular} 
\caption{\label{tab.hans}Hansen coefficients $X_0^{n,m}$ for $(0\leq n \leq 10, 0\leq m \leq n)$.}
\end{table}
}

\newcommand{\tabd}{
\begin{table}[h!]
\begin{tabular}{|cc| l | cc| l|}
\hline
\SSa $n$ &$m$ &\hfil $\tilde X_0^{-n,m}$ & $n$ &$m$ &\hfil $\tilde X_0^{-n,m}$\\
\hline
\SSb$2$ & $0$ & $1$ &                                                        $8$ & $0$ & $ +1+ {15}/{2}\,e^{2}+ {45}/{8}\,e^{4}+ {5}/{16}\,e^{6}$\\                     
\SSb$3$ & $0$ & $1$ &                                                                $8$ & $1$ & $ +3\,e+ {15}/{2}\,e^{3}+ {15}/{8}\,e^{5}$\\                              
\SSb$3$ & $1$ & $ + {1}/{2}\,e$ &                                            $8$ & $2$ & $ + {15}/{4}\,e^{2}+ {15}/{4}\,e^{4}+ {15}/{64}\,e^{6}$\\                     
\SSb$4$ & $0$ & $ +1+ {1}/{2}\,e^{2}$ &                                      $8$ & $3$ & $ + {5}/{2}\,e^{3}+ {15}/{16}\,e^{5}$\\                     
\SSb$4$ & $1$ & $ +\,e$ &                                                   $8$ & $4$ & $ + {15}/{16}\,e^{4}+ {3}/{32}\,e^{6}$\\                                  
\SSb$4$ & $2$ & $ + {1}/{4}\,e^{2}$ &                                       $8$ & $5$ & $ + {3}/{16}\,e^{5}$\\                                        
\SSb$5$ & $0$ & $ +1+ {3}/{2}\,e^{2}$ &                                        $8$ & $6$ & $ + {1}/{64}\,e^{6}$\\                       
\SSb$5$ & $1$ & $ + {3}/{2}\,e+ {3}/{8}\,e^{3}$ &                           $9$ & $0$ & $ +1+ {21}/{2}\,e^{2}+ {105}/{8}\,e^{4}+ {35}/{16}\,e^{6}$\\                                      
\SSb$5$ & $2$ & $ + {3}/{4}\,e^{2}$ &                                       $9$ & $1$ & $ + {7}/{2}\,e+ {105}/{8}\,e^{3}+ {105}/{16}\,e^{5}+ {35}/{128}\,e^{7}$\\                                        
\SSb$5$ & $3$ & $ + {1}/{8}\,e^{3}$ &                                      $9$ & $2$ & $ + {21}/{4}\,e^{2}+ {35}/{4}\,e^{4}+ {105}/{64}\,e^{6}$\\                                                   
\SSb$6$ & $0$ & $ +1+3\,e^{2}+ {3}/{8}\,e^{4}$ &                            $9$ & $3$ & $ + {35}/{8}\,e^{3}+ {105}/{32}\,e^{5}+ {21}/{128}\,e^{7}$\\                                      
\SSb$6$ & $1$ & $ +2e+ {3}/{2}\,e^{3}$ &                                    $9$ & $4$ & $ + {35}/{16}\,e^{4}+ {21}/{32}\,e^{6}$\\                                      
\SSb$6$ & $2$ & $ + {3}/{2}\,e^{2}+ {1}/{4}\,e^{4}$ &                         $9$ & $5$ & $ + {21}/{32}\,e^{5}+ {7}/{128}\,e^{7}$\\                                               
\SSb$6$ & $3$ & $ + {1}/{2}\,e^{3}$ &                                       $9$ & $6$ & $ + {7}/{64}\,e^{6}$\\                                         
\SSb$6$ & $4$ & $ + {1}/{16}\,e^{4}$ &                                     $9$ & $7$ & $ + {1}/{128}\,e^{7}$\\                                                       
\SSb$7$ & $0$ & $ +1+5e^{2}+ {15}/{8}\,e^{4}$ &                             $10$ & $0$ & $ +1+14\,e^{2}+ {105}/{4}\,e^{4}+ {35}/{4}\,e^{6}+ {35}/{128}\,e^{8}$\\                                      
\SSb$7$ & $1$ & $ + {5}/{2}\,e+ {15}/{4}\,e^{3}+ {5}/{16}\,e^{5}$ &         $10$ & $1$ & $ +4\,e+21e^{3}+ {35}/{2}\,e^{5}+ {35}/{16}\,e^{7}$\\                                       
\SSb$7$ & $2$ & $ + {5}/{2}\,e^{2}+ {5}/{4}\,e^{4}$ &                       $10$ & $2$ & $ +7\,e^{2}+ {35}/{2}\,e^{4}+ {105}/{16}\,e^{6}+ {7}/{32}\,e^{8}$\\                                                
\SSb$7$ & $3$ & $ + {5}/{4}\,e^{3}+ {5}/{32}\,e^{5}$ &                    $10$ & $3$ & $ +7e^{3}+ {35}/{4}\,e^{5}+ {21}/{16}\,e^{7}$\\                                                                      
\SSb$7$ & $4$ & $ + {5}/{16}\,e^{4}$ &                                     $10$ & $4$ & $ + {35}/{8}\,e^{4}+ {21}/{8}\,e^{6}+ {7}/{64}\,e^{8}$\\                                                       
\SSb$7$ & $5$ & $ + {1}/{32}\,e^{5}$ &                                     $10$ & $5$ & $ + {7}/{4}\,e^{5}+ {7}/{16}\,e^{7}$\\                                                        
\SSb&& &                                     $10$ & $6$ & $ + {7}/{16}\,e^{6}+ {1}/{32}\,e^{8}$\\                                       
\SSb&& &                                     $10$ & $7$ & $ + {1}/{16}\,e^{7}$\\                                       
\SSc&&&                                                                     $10$ & $8$ & $ + {1}/{256}\,e^{8}$\\ 
\hline
\end{tabular} 
\caption{\label{tab.hansn}$\tilde X_0^{-n,m}$ is the polynomial part of the Hansen coefficients  $X_0^{-n,m}$. For $(0\leq n \leq 10, 0\leq m \leq n-2)$, 
the full Hansen coefficients
are $X_0^{-n,m}= \tilde X_0^{-n,m} / (1-e^2)^{n-3/2}$. For $n\geq 2$, we have also $X_0^{-n,n-1}=X_0^{-n,n}=0$.
For $n=1$, we have   $X_0^{-1,0}=1$ and $X_0^{-1,1}=(\sqrt{1-e^2}-1)/e$.   }
\end{table}
}

\begin{document}

\numberwithin{equation}{section}

\title{Explicit expansion of the three-body disturbing function for arbitrary eccentricities and  inclinations}
\author{Jacques Laskar \and Gwena\"el Bou\'e}
\titlerunning{Disturbing function for arbitrary eccentricities and  inclinations}
\authorrunning{J. Laskar \and G. Bou\'e}
\date{\today}
\institute{ASD, IMCCE-CNRS UMR8028, Observatoire de Paris, UPMC, 
           77 avenue Denfert-Rochereau, 75014 Paris, France}
\abstract{Since the original work of Hansen and Tisserand in the XIXth century, there 
have been many variations in the analytical expansion of the three-body disturbing function 
in  series of the  semi-major axis  ratio. With the increasing number of planetary systems 
of large eccentricity,  these expansions are even more interesting as they allow us
to obtain for the secular systems finite expressions that are valid for all eccentricities 
and inclinations. 
We revisited the derivation of the disturbing function in Legendre polynomial, 
with a special focus on the secular system. We provide here  expressions 
of the disturbing function for the planar and spatial case at any order 
with respect to the ratio of the semi-major axes. Moreover, for   orders in the 
ratio of semi-major axis up to  ten  
in the planar case and five in the spatial case, we provide explicit expansions 
of the secular system, and simple algorithms with minimal computation to extend 
this to higher order, as well as the algorithms for the computation of non secular terms.
}

\maketitle

\section{Introduction}
In the three-body problem, there are two classical ways to compute the principal part of the disturbing function 
 $-\cG mm'/\Delta$. The first approach is to expand it in powers of eccentricity and inclination, 
 with coefficients that are expressed in term of Laplace coefficients
 \citep[e.g.~][]{Laplace_1785,abu-el-ata_analytical_1975,laskar_stability_1995}, 
 but this approach, which is well suited to the study of the Solar  System, 
 has some limitations for some extra-solar planetary systems,  where the eccentricity can reach very high values. 
 Another drawback, is that without some special truncation 
 corrections, the angular momentum will not  be  conserved exactly in the truncated system.
 
 In the second approach, the expansion is made with respect to the ratio of the semi-major axes 
of the two bodies  $\alpha = a/a'$, where $a'$ is related to the external body. 
Although it may be less efficient for low eccentricity and large values of 
$\alpha$, as for the inner planets in the Solar system, the    advantage is that 
this expansion allows us to obtain finite expressions for arbitrary eccentricities
in the secular system, while both approaches allow expansions for arbitrary inclinations.
The most important contribution to this problem was  made in the XIXth century 
\citep{Hansen_1855,hill_development_1875,Tisserand_1899}. With the development of 
space technology,  expansions  of the disturbing function 
in term of Legendre polynomials have been rejuvenated for the construction of 
satellite theories in the vicinity of the Earth 
\citep{kozai_motion_1959,kaula_development_1962,Brumberg:1967:a,brumberg_analytical_1971,giacaglia_lunar_1974,abu-el-ata_analytical_1975}.

The discovery of new planetary systems, starting with 51Pegb \citep{mayor_jupiter-mass_1995}, 
has rasemi-majorised the need to revise these methods as many systems have planets with very high eccentricity, 
as  GL581, HD217107, HD69830, HD74156, HD168443, HD102272, HD169830, HD202206, HD183263,
or even HD80606, where the  eccentricity reaches 0.93\footnote{see  http://exoplanet.eu}.
With the discovery of these numerous new planetary  systems, the  previous analytical expansions
in Laplace coefficients developed for the Solar System are no longer the most appropriate, 
and we are  faced with the  need to understand more globally the dynamics of these 
systems. On the other hand, as the parameters of these extrasolar planetary systems remain not very well known, 
there is no necessity for very precise analytical approximations. 
This has led several authors to use the Legendre expansion of the potential for the 
dynamical study of the secular planetary system, following the previous studies 
of stellar systems \citep{krymolowski_studies_1999,ford_secular_2000,blaes_kozai_2002},
where the secular spatial three-body system was expanded up to the octupolar order ($\al^3$).
In particular, \citet{lee_secular_2003}
 used the secular planar system at octupolar order to study the secular dynamics of the 
HD168443 and HD12661 systems.
\citet{migaszewski_secular_2008}  computed the planar secular system 
to high order using computer algebra  to average over the mean anomaly. 

As it appears that there is a growing interest in analytical studies of extra solar planetary systems, 
we considered it  interesting to present a derivation of the planetary (or stellar) 
disturbing function in a very simple and explicit way that does not already appear  
 in the existing literature. We  aimed to write a self-contained paper that allows one
to construct explicitly the planetary Hamiltonian to high order, with minimal 
additional computation. 
In particular, in Sect. \ref{sec_plan} we show that the planar secular system can be obtained explicitly at any order without 
the need for computer algebra. The spatial case is treated in Sect. \ref{sec_spa} 
with minimal computation when expressed  with respect to the mutual inclination $J$. 
In the presence  of more than two planets, it may be more convenient to use 
a fixed reference frame, and the  derivation of the spatial  Hamiltonian 
is provided in this case in Sect. \ref{sec_spaii}.
Our derivation is  close to the original formulations of \citet{Hansen_1855} and \citet{Tisserand_1899}, but with 
emphasis on  the secular system and the direct derivation of 
explicit expressions. We also present  in the
Appendix \ref{an_hans} a  new   derivation of the Hansen coefficients for the 
secular terms.

\section{Expansion of the disturbing function}
To simplify the notations, and although everything can be generalized to 
an $N$-body system, we  consider here a three-body problem 
with a central body of mass $M$ and two other celestial bodies of masses $m,m'$.
If we write the Hamiltonian of the Newtonian interactions among these 
three bodies in Poincar\'e canonical heliocentric variables, 
we obtain \citep[][Eq. 15]{laskar_stability_1995}
\be
{\cal H} = {\cal H}_0 + {\cal T}_1 + {\cal U}_1 \ ,
\ee
where 
\be
{\cal H}_0 = -\Frac{G(M+m)}{2a} -\Frac{G(M+m')}{2a'} 
\ee
is the Keplerian interaction  
with  elliptical elements  $a,e,i,M, \om$, and $\Om$ that denote, respectively, the 
semi-major axis, eccentricity, inclination, mean anomaly, argument of perihelion, 
and longitude of the node (with primes for the external body of mass $m'$).
The principal part 
${\cal U}_1$  
part of the perturbation and indirect part ${\cal T}_1$ are then 
\be
{\cal U}_1 = - G \Frac{mm'}{\norm{\br-\br'}} \ ; \quad {\cal T}_1 = \Frac{\tilde\br\cdot \tilde\br'}{M} \ ,
\label{eq.hamt}
\ee
where $\br, \br'$ are  the radius vectors of the inner and outer planets,
with norms
$r, r'$,  unit vectors $\uu= \br/r$ and $\uu'=\br'/r'$,
and conjugate momentum $\tilde\br, \tilde\br'$.
We  focus first on the  principal part of the Hamiltonian ${\cal U}_1$, 
which is the most difficult part to compute, while the
computation of the indirect part will be made  in Sect. \ref{sec.comp}.
With $\rho=r/r'$, and $\cF = a'/\norm{\br-\br'}$, we have
\be
\cF= \frac{a'}{r'}\left(1+\rho^2 -2\rho \uu\dpp\uu'\right)^{-1/2} = 
\frac{a' }{r' }\sum_{n=0}^\infty   P_n(\uu\dpp\uu') \rho^n\ ,
\ee
where $P_n(x)$ are the Legendre polynomials that can be written as 
\be
P_n(z) = 
\sum_{k=0}^{[n/2]} p_{n,k}\, z^{n-2k}
\ee
with
\be
p_{n,k} = \frac{(-1)^k}{2^n} \frac{(2n-2k)!}{k! (n-k)! (n-2k)!} \ .
\label{eq.pnk}
\ee

With $\al=a/a', \ga=r/a, \ga'=r'/a'$, we have $\rho=\al \ga /\ga'$, 
and thus, as $\ga,\ga',\uu, \uu'$
do not depend on $a,a'$, the expansion of $\cF$ in powers of $\al$ is 

\be
\cF = \sum_{n=0}^\infty  \cF_n \, \al^n\ ;\quad \hbox{with}\ \cF_n=  \Frac{\ga^n}{\ga'^{n+1}} F_n
\ee
and
\be
 F_n=P_n(\uu\dpp\uu')=\sum_{k=0}^{[n/2]} p_{n,k}\, (\uu\dpp\uu')^{n-2k}
\label{eq.fn}
\ee

\section{Planar case}
\label{sec_plan}
\begin{table*}
\begin{tabular}{r l}
\hline\hline
$F_{0} $&= $1 $\\[5pt]
$F_{1} $&= $\cos x $\\[5pt]
$F_{2} $&= $(1+3\cos 2x)/4$ \\[5pt]
$F_{3} $&= $(3\cos x  +5\cos 3x)/8 $\\[5pt]
$F_{4} $&= $(9 + 20\cos 2x  + 35\cos 4x)/64 $\\[5pt]
$F_{5} $&= $(30\cos x + 35\cos 3x + 63\cos 5x)/128 $\\[5pt]
$F_{6} $&= $(50+ 105\cos 2x + 126\cos 4x + 231\cos 6x)/512 $\\[5pt]
$F_{7} $&= $(175\cos x + 189\cos 3x + 231\cos 5x + 429\cos 7x)/1024 $\\[5pt]
$F_{8} $&= $(1225+ 2520\cos 2x + 2772\cos 4x + 3432\cos 6x + 6435\cos 8x)/16384 $\\[5pt]
$F_{9} $&= $(4410\cos x + 4620\cos 3x + 5148\cos 5x + 6435\cos 7x + 12155\cos 9x)/32768 $\\[5pt]
$F_{10}$&= $ (7938 + 16170\cos 2x  + 17160\cos 4x + 19305\cos 6x + 24310\cos 8x + 46189\cos 10x)/131072 $\\[5pt]
\hline
\end{tabular} 
\caption{\label{tabfnfn}Tisserand functions for the planar case (Eq.\ref{eq.fn}). We have $x=u-u'$ with $u=v+\om, u'=v'+\om'$.}
\end{table*}

\tabb

We  first study the planar case that leads to some simplifications in the expansions. 
We define 
 $v,v'$ to be the true anomalies of $\uu,\uu'$, and $\om, \om'$ their argument of perihelion,  
 $u=v+\om$, $u'=v'+\om'$, and $x=u-u'$. 
 \mybf{
 We have then 
  $\uu\dpp\uu' = \cos x $ and, following a classical computation \citep[e.g.][p. 303]{Whittaker:1927:a}, we
  have for all $z \in [0,1[$, 
\be
\EQM{
(1-2z\cos x+z^2)^{-1/2}  =  
(1-z \,\e^{\i x})^{-1/2}(1-z \,\e^{-\i x})^{-1/2} \crm
= \sum_{p=0}^{\infty}\sum_{q=0}^{\infty}\frac{(2p)!(2q)!}{2^{2p+2q}(p!)^2(q!)^2} \,\e^{\i(p-q)x} z^{p+q}
}
\ee
and, with $n=p+q$, and  after changing $q$ to $n-q$,
\be
= \sum_{n=0}^{\infty}
\left(\sum_{q=0}^n \frac{(2q)!(2n-2q)!}{2^{2n}(q!)^2((n-q)!)^2}
\,\e^{\i(2q-n)x}\right)z^n.
\ee
Thus
\be
F_n=  \sum_{q=0}^{n} f_{n,q}  \e^{\i(2q-n)x} 
\label{eq.fnx}
\ee
with 
\be
f_{n,q}=  \frac{(2q)!(2n-2q)!}{2^{2n}(q!)^2((n-q)!)^2}  
\label{eq.fnq}
\ee
for $0\leq q \leq n$.
}
If we write $x = v-v'+\om-\om'$, 
$\cF_n$ can now be expressed  in the form 

\be
\cF_n=  \sum_{q=0}^{n} \cF_{n,q} \, \e^{\i(2q-n)(\om-\om')} 
\ee
with 
\be
 \cF_{n,q} = f_{n,q}  \Frac{\ga^n}{\ga'^{n+1}} \e^{\i(2q-n)(v-v')}
\ee
The quantities $\cF_{n,q}$ can then be expressed in term of Hansen coefficients $X_k^{n,m}(e)$ defined for $n,m \in \mathbb{Z}$ as 
\be
\left(\frac{r}{a}\right)^n \e^{\i mv}  = \sum_{k=-\infty}^{+\infty} X_k^{n,m}(e) \, \e^{\i kM} \ .
\ee
For convenience, we  denote $X_k^{n,m}=X_k^{n,m}(e)$ and ${X'}_k^{n,m}=X_k^{n,m}(e')$.
We have thus 
\be
 \cF_{n,q} = f_{n,q} \sum_{k,k'=-\infty}^{+\infty} X_k^{n,2q-n} {X'}_{k'}^{-(n+1),-2q+n}\e^{\i(kM+k'M')}\ .
\ee
For arbitrary values of $k\in \mathbb{Z}$, the  Hansen coefficients  $X_k^{n,m}$  can be expressed 
in an explicit manner as an infinite  series involving Bessel functions and 
 hypergeometric functions \citep{Hansen_1855,Tisserand_1899}, or in terms of generalized Laplace coefficients
\citep{laskar_notegeneralized_2005}, 
but for $k = 0$, $X_k^{n,m}$ reduces to  a finite polynomial expression in $e$, $1/e$, $\sqrt{1-e^2}$, and 
$1/\sqrt{1-e^2}$ (see Appendix A). 
We have thus a very  compact expression for the coefficient of any argument $kM+k'M'$ 
in explicit form at all orders $n$ in $\alpha$. 
Indeed, if we denote this coefficient  $\cF_n^{(k,k')}$,  that is 
\be
\cF_n =  \sum_{k,k'=-\infty}^{+\infty} \cF_n^{(k,k')} \e^{\i(kM+k'M')} \ ,
\label{eq.fkk}
\ee
we have
\be
\cF_n^{(k,k')} = \sum_{q=0}^{n} f_{n,q} X_k^{n,2q-n}  {X'}_{k'}^{-(n+1), -2q+n} \e^{\i(2q-n)(\om-\om')}  \ .
\ee

In particular, for all $n\in \mathbb{N}$, the secular part $\cF_n^{(0,0)}$ can even be simplified, 
 using  the classical relation $X_{-k}^{n,-m}= X_k^{n,m}$ among Hansen coefficients
 and the relation $f_{n,q}=f_{n,n-q}$ from Eq.(\ref{eq.fnq})
\be
\EQM{
\cF_n^{(0,0)} = \epsilon_n f_{n,\frac{n}{2}} X_0^{n,0}  {X'}_{0}^{-(n+1), 0}  + \crm
\sum_{q=0}^{[(n-1)/2]} 2\,f_{n,q} X_0^{n,n-2q}  {X'}_{0}^{-(n+1), n-2q}  \cos((n-2q)(\om-\om') ) \ ,
}
\label{eq.perplan}
\ee
where $\epsilon_n = 0$ if $n$ is odd, and $\epsilon_n = 1$ if $n$ is even.

\subsection{Practical algorithm}

Equations (\ref{eq.fnq}) and (\ref{eq.perplan}) provide an explicit algorithm for the computation 
of the principal part of  the disturbing function $\cF$, and in particular for the computation of the secular 
Hamiltonian at any order $N$ in $\alpha$
\be
<\cF> = \sum_{n=0}^N \cF_n^{(0,0)}\alpha^n \ ,
\ee
for which the involved Hansen coefficients $ X_0^{n,m},  {X'}_{0}^{-(n+1), m}$ reduce to finite expressions 
(see Appendix A).
As there are no secular terms in the indirect part of the Hamiltonian 
\citep[see][]{laskar_stability_1995}, the computation
of the secular Hamiltonian reduces to the computation of the rational constants $f_{n,k}$ 
given by Eq.(\ref{eq.fnq}). 
In practice, for finite order $n$, it is even easier to use the expression of $F_n$ 
(\ref{eq.fnx}) (Table \ref{tabfnfn}) and to 
translate it into $\cF_n$  (\ref{eq.perplan}) by the simple transformation
\be
\EQM{
 \Frac{\ga^n}{\ga'^{n+1}}\cos (mx) =\crm \sum_{{k,k'=-\infty}}^{+\infty} 
  X_k^{n,m}  {X'}_{k'}^{-(n+1), -m} \cos(kM+k'M'+m(\om-\om') )   \ ,
  }
\label{eq.transf}
\ee
which allows an 
immediate computation of any argument 
$\cF_n^{(k,k')}$ in terms of Hansen coefficients.

\subsection{Computation of the secular part $\cF_n^{(0,0)}$ }

The computation of the secular part of the Hamiltonian $\cF_n^{(0,0)}$ is quite simple, 
as we will just have to make the transformation 
\be
\cos (mx) \longrightarrow  
 X_0^{n,m} {X'}_{0}^{-(n+1), m}  \cos(m(\om-\om') ) 
\ee
in the expression of $F_n$. Moreover, we have ${X'}_{0}^{-(n+1), m}= 0$ for $m \geq n \geq 1$ (see Appendix \ref{an_hans}). The term in $\cos n x$ 
can thus be discarded in the expression of $F_n$ (Table \ref{tabfnfn})
which simplifies   the expression of the Hamiltonian. The secular Hamiltonian is thus expressed in finite 
form, using the values  of the Hansen coefficients given in Appendix
\ref{an_hans}.
We have
$$
\EQM{
\cF_0^{(0,0)} &= 1\cr
\cF_1^{(0,0)} &= 0\cr
\cF_2^{(0,0)} &= \Frac{1}{4} X_0^{2,0}{X'}_{0}^{-3, 0}\cr
\cF_3^{(0,0)} &= \Frac{3}{8} X_0^{3,1}   {X'}_{0}^{-4, 1}   \cos(\om-\om' )\cr
\cF_4^{(0,0)} &= \Frac{9}{64} X_0^{4,0}  {X'}_{0}^{-5, 0}  + \Frac{5}{16} X_0^{4,2}  {X'}_{0}^{-5, 2}  \cos(2(\om-\om') )\cr
\cF_5^{(0,0)} &= \Frac{15}{64} X_0^{5,1}   {X'}_{0}^{-6, 1}   \cos( \om-\om')\cr
              &+ \Frac{35}{128} X_0^{5,3}   {X'}_{0}^{-6, 3}   \cos(3(\om-\om') )  \cr
\cF_6^{(0,0)} &= \Frac{25}{256}X_0^{6,0}   {X'}_{0}^{-7, 0} +\Frac{105}{512} X_0^{6,2}   {X'}_{0}^{-7, 2}  \cos( 2(\om-\om'))\cr
              &+ \Frac{63}{256} X_0^{6,4}   {X'}_{0}^{-7, 4}   \cos(4(\om-\om') )\cr                   
\cF_7^{(0,0)} &= \Frac{175}{1024} X_0^{7,1}   {X'}_{0}^{-8, 1} \cos(\om-\om')\cr
               &+ \Frac{189}{1024} X_0^{7,3}   {X'}_{0}^{-8, 3} \cos(3(\om-\om'))\cr
               &+ \Frac{231}{1024} X_0^{7,5}   {X'}_{0}^{-8, 5} \cos(5(\om-\om'))\ .
}
$$
These results are equivalent to the expression obtained by \citet{migaszewski_secular_2008} using 
computer algebra. 
\section{Spatial case}
\label{sec_spa}

The spatial case is more complicated as it involves additional variables. 
Our goal  is to derive explicit  formulae that are as compact as possible. We  thus 
expand in terms of the mutual inclination $J$. 
For each orbit, we use  a reference frame $(\bi,\bj,\bk)$ associated with  the orbit, with first vector $\bi$ in the direction of the 
ascending node of $\br'$ over $\br$. With $u=v+\om$ and $u'=v'+\om'$, we have 
\be
\EQM{
\uu\dpp\uu' &= \cos u \cos u' + \cos J \sin u \sin u' \cr
            &= \mu \cos x + \nu \cos y
\label{eq.uupJ}            
}
\ee
with the same notations as \citet{tisserand_memoire_1885}
\be
x = u-u' \ ;\quad y= u+u' \ ;\quad  \mu = \cos^2 \Frac{J}{2} \ ;\quad  \nu = \sin^2 \Frac{J}{2} \ .
\ee
As in the planar case, we  have   $F_n=P_n(\uu\dpp\uu')$, but now $\uu\dpp\uu'$ is given by the
slightly more complex expression  (\ref{eq.uupJ}). For all $n$, we have 
\be
F_n =  
\sum_{s=0}^{n} \sum_{q= 0}^{n }  \cQ^{(n)}_{s,q} (\mu,\nu) \, \e^{\i u\,(n-2s)}\e^{\i u'\,(n-2q)} \ ,
\label{eq.fnuu}
\ee
\mybf{where the $ \cQ^{(n)}_{s,q} (\mu,\nu)$ are polynomials in $\mu,\nu$ of degree $n$ that are called the 
Tisserand polynomials\footnote{One should consult \citep{aksenov_tisserands_1986} for a detailed discussion of the 
relation between the Tisserand polynomials and the inclination functions of \citet{kaula_development_1962}.}
as a recognition of the work of \citet{tisserand_memoire_1885},
although these expressions are already present in \citep{Hansen_1855}.}
As $(u,u') \longrightarrow (-u, -u')$ leaves $\uu\dpp\uu'$ unchanged, we have 
$ \cQ^{(n)}_{n-s,n-q} = \cQ^{(n)}_{s,q} $,  thus $F_n$ can be expressed as a trigonometric polynomial 
in $\cos (mu+m'u')$. 
Although they can be computed explicitly for all $n$ (see Appendix B), for  a given $n$, it is 
often more efficient to make a direct computation 
of $F_n$ on a computer algebra system.  For example, $F_{20}$ is computed in 
less than $1.5$ seconds in exact rational arithmetics with TRIP \citep{GL09} on an average laptop computer using 
the simple expression (\ref{eq.fn}). 
We can then express $\cF_n = (\ga^n/\ga'^{n+1}) F_n$ in term of Hansen coefficients. We thus have 
\be
\cF_n =  \sum_{k=-\infty}^{+\infty}\sum_{k'=-\infty}^{+\infty} \cF_n^{(k,k')} \e^{\i(kM+k'M')} \ ,
\ee
with  
\be
\EQM{
\cF_n^{(k,k')} =\crm
\sum_{s,q=0}^{n}   \cQ^{(n)}_{s,q}
\,  X_k^{n,n-2s}\,  {X'}_{k'}^{-(n+1),n-2q}\,\e^{\i\,(n-2s)\om}\e^{\i\,(n-2q)\om'} \ .
}
\ee
More practically, starting from the finite expressions in cosine polynomials 
(Table \ref{eq.fnmunuii}) of the Tisserand functions $F_n$, 
the 
expression of $\cF_n$ is obtained  as in the planar case by means of the more general  transformation 
\be
\EQM{
 \Frac{\ga^n}{\ga'^{n+1}}\cos (mu+m'u')=\sum_{k=-\infty}^{+\infty}\sum_{{k'=-\infty}}^{+\infty}\crm
  X_k^{n,m}  {X'}_{k'}^{-(n+1), m'}  \cos(kM+k'M'+m\om+m'\om')  \ .
 }
\label{eq.transfiiii}
\ee

\subsection{Computation of the secular part $\cF_n^{(0,0)}$ }

As in the planar case, the computation of 
 the secular part of the Hamiltonian $\cF_n^{(0,0)}$ is just a straightforward translation 
 of $F_n$ in Table \ref{eq.fnmunuii}, with the transformation 
 \be
 \cos (mu+m'u') \longrightarrow X_0^{n,m} {X'}_{0}^{-(n+1), m'}  \cos(m\om+m'\om') \ .
 \ee
Moreover, as ${X'}_0^{-(n+1),n} = 0$ for $n \geq 1$, all terms in $\cos(mu\pm nu')$ can be discarded in $F_n$. 
We thus have, 
$$
\EQM{
\cF_{0}^{(0,0)} =1\cr
\cF_{1}^{(0,0)} =0\cr
\cF_{2}^{(0,0)} =  \cr
 (- \Frac{1}{2}
 + \Frac{3}{4} \nu^2
 + \Frac{3}{4} \mu^2)X_0^{2,0}{X'}_{0}^{-3, 0} +  
  \Frac{3}{2} \nu \mu X_0^{2,2}{X'}_{0}^{-3, 0} \cos(2\om) \crm
\cF_{3}^{(0,0)} =\cr
X_0^{3,1} {X'}_{0}^{-4,1} \left[ (- \frac{3}{2} \mu  + \frac{15}{4} \nu^2 \mu   + \frac{15}{8} \mu^3 )   \cos(\om-\om') \right.
 \crm
 +\left.(- \frac{3}{2} \nu    + \frac{15}{8} \nu^3   + \frac{15}{4} \nu \mu^2)  \cos(\om+\om')\right] \crm
 +\frac{15}{8}X_0^{3,3}  {X'}_{0}^{-4,1} \left[  \nu^2 \mu  \cos(3\om+\om') 
 +  \nu \mu^2    \cos(3\om-\om')\right]\crm
 }
$$
$$
\EQM{
\cF_{4}^{(0,0)} =X_0^{4,0}{X'}_{0}^{-5, 0}  \crm
\left(
 + \Frac{3}{8}
 - \Frac{15}{8} \nu^2
 + \Frac{105}{64} \nu^4
 - \Frac{15}{8} \mu^2
 + \Frac{105}{16} \nu^2 \mu^2
 + \Frac{105}{64} \mu^4\right) \crm
  + X_0^{4,2}{X'}_{0}^{-5, 2}\crm
  \left[(
 - \Frac{15}{8} \mu^2  
 + \Frac{105}{16} \nu^2 \mu^2  
 + \Frac{35}{16} \mu^4) \cos(2\om-2\om') \right.\crm
 +\left.(
 - \Frac{15}{8} \nu^2  
 + \Frac{35}{16} \nu^4  
 + \Frac{105}{16} \nu^2 \mu^2 )\cos(2\om+2\om')\right]\crm
+(
 - \Frac{15}{4} \nu \mu 
 + \Frac{105}{16} \nu^3 \mu  
 + \Frac{105}{16} \nu \mu^3) X_0^{4,2}{X'}_{0}^{-5, 0}\cos(2\om) \crm
 +(
 - \Frac{15}{4} \nu \mu  
 + \Frac{105}{16} \nu^3 \mu  
 + \Frac{105}{16} \nu \mu^3) X_0^{4,0}{X'}_{0}^{-5, 2} \cos(2\om')\crm
+X_0^{4,4}{X'}_{0}^{-5,2}\crm
\left[
 + \Frac{35}{16} \nu \mu^3 \cos(4\om-2\om')
 + \Frac{35}{16} \nu^3 \mu \cos(4\om+2\om')\right]\crm
+ \Frac{105}{32}X_0^{4,4}{X'}_{0}^{-5, 0} \nu^2 \mu^2 \cos(4\om)\ .
}
$$
We thus observe here an important simplification in the quadrupolar  secular Hamiltonian $\cF_{2}^{(0,0)}$, as all terms 
involving the external planet longitude of perihelion $\om'$ vanish and we are left with an integrable Hamiltonian. This is what 
\citet{lidov_non-restricted_1976}  called a happy coincidence 
\citep[see also][]{farago_high-inclination_2010}. This is no longer the case at higher orders.

\newcommand{\co}{{\rm c}}
\newcommand{\si}{{\rm s}}

\subsection{Expression in a fixed reference frame}
\label{sec_spaii}
\tabii

The above expressions are given with respect to the mutual inclination  to shorten the algebraic 
expansions. This is especially useful in a three-body problem, but
 to obtain expressions valid in a fixed reference frame, then one needs to substitute into 
$\uu\dpp\uu'$ its expression in terms of the elliptical elements of the two bodies. We can then 
generalize the expression (\ref{eq.uupJ}) and write it now as \citep{Brumberg:1967:a,abu-el-ata_analytical_1975,laskar_stability_1995} 
\be
\uu\dpp\uu' = \Re \left( \tmu \,\e^{\i x} + \tnu \,\e^{\i y}\right)
\ee
with as before $x=u-u'$, $y=u+u'$, and 
\be
\EQM{
\tmu &=  \left( \co \co'\e^{\i\frac{\Om-\Om'}{2}} + \si\si' \e^{-\i\frac{\Om-\Om'}{2}}  \right)^2 \crm
\tnu &=  \left( \co \si'\e^{\i\frac{\Om-\Om'}{2}} - \si\co' \e^{-\i\frac{\Om-\Om'}{2}}  \right)^2
}
\ee
with $\co = \cos(i/2)$, $\si = \sin(i/2)$, and the same for the primes. We write
\be
\tmu = a + i a' \ ; \quad \tnu = b+ib' 
\ee
with
\be
\EQM{
a &= (\co^2{\co'}^2+\si^2{\si'}^2) \cos (\Om-\Om') + 2 \co\co'\si\si' \cr
a' &= (\co^2{\co'}^2-\si^2{\si'}^2) \sin (\Om-\Om')  \cr
b &= (\co^2{\si'}^2+\si^2{\co'}^2) \cos (\Om-\Om') - 2 \co\co'\si\si' \cr
b' &= (\co^2{\si'}^2-\si^2{\co'}^2) \sin (\Om-\Om') \ . 
}
\label{eq.aabb}
\ee
\mybf{
With these notations, we have 
\be
\abs{\tnu} = \sin^2(J/2), \qquad
\abs{\tmu} = \cos^2(J/2)
\label{eq.tnutmu}
\ee
and
}
\be
\EQM{
\uu\dpp\uu' &= &a\cos(u-u') -a'\sin(u-u')\cr &+&b\cos(u+u')-b'\sin(u+u') \ .
\label{eq.uuii}
}
\ee
The remaining part is identical to the previous case.
The   expressions of the  
Tisserand functions $F_n$ are given for $0\leq n\leq 3$ in Table \ref{eq.fnmunu} and for all $n$ in  Appendix
\ref{sec.func_tisserand}. 
In practice, for small  values of $n$ we can use  explicit expressions of $F_n$,  
using the straightforward approach consisting of computing $F_n$ directly 
with computer algebra, using the relation (\ref{eq.fn}), and then to translate it as previously in order to obtain 
the expression of any argument $kM+k'M'$ by means of the relations
\be
\EQM{
 \Frac{\ga^n}{\ga'^{n+1}}\left\{\EQM{\cos\cr\sin}\right. ( m u+m'u')=\sum_{k=-\infty}^{+\infty}\sum_{{k'=-\infty}}^{+\infty}\crm
  X_k^{n,m}  {X'}_{k'}^{-(n+1), m'}  \left\{\EQM{\cos\cr\sin}\right.(kM+k'M'+m\om+m'\om')  \ .
 }
\label{eq.transfii}
\ee
We thus have for the secular part $\cF_n^{(0,0)}$,
$$
\EQM{
\cF_{0}^{(0,0)} =1\cr
\cF_{1}^{(0,0)} =0\cr
\cF_{2}^{(0,0)} =  X_0^{2,0}{X'}_{0}^{-3, 0}[
 -\Frac{1}{2}
 +\Frac{3}{4}( b'^2
 + b^2
 + a'^2
 + a^2) ]\crm
+ X_0^{2,2}{X'}_{0}^{-3, 0}\, \Frac{3}{2}[ 
 ( b a
 -  b' a') \cos(2 \om)
 -(  b a'  
 +  b' a )\sin(2 \om) ]
}
$$
and
$$
\EQM{
\cF_{3}^{(0,0)}= 
 \Frac{15}{8}X_0^{3,1}{X'}_{0}^{-4, 1}[\crm
 +(2 b'^2 a  +2 b^2 a    + a'^2 a  + a^3     -\Frac{4}{5} a      ) \cos(\om-\om')\crm
 +(2 b a'^2   +2 b a^2    -\Frac{4}{5} b         +  b'^2 b   +  b^3    )\cos(\om+\om')\crm
 +(\Frac{4}{5} b'    -  b'^3    -  b' b^2  -2 b' a'^2 -2b' a^2  )\sin(\om+\om')\crm
 -(  a' a^2  -\Frac{4}{5} a'        +2 b'^2 a'  +2 b^2 a'   +  a'^3  )  \sin(\om-\om')]\crm 
 +  \Frac{15}{8}X_0^{3,3}{X'}_{0}^{-4, 1}[\crm
 ( b^2 a  -2 b' b a'   - b'^2 a )\cos(3 \om+\om')\crm
 +(b a^2   - b a'^2   -2 b' a' a )\cos(3 \om-\om')\crm
 +( b'^2 a'   - b^2 a'   -2 b' b a )\sin(3 \om+\om')\crm
 -(2 b a' a   +  b' a^2   -  b' a'^2 )\sin(3 \om-\om')] \ .
}
$$

\section{Indirect part}
\label{sec.comp}
Until  now, we  considered only the principal part of the perturbing Hamiltonian. The computation of
the indirect part ${\cal T}_1$ (\ref{eq.hamt}) is more straightforward, as it compares with the computation of 
$\br\dpp\br'$. Indeed, we have  \citep[][Eq. 24]{laskar_stability_1995},
\be
{\cal T}_1 = \Frac{mm'}{M} V \quad \hbox{with} \quad V=\dot\br\dpp\dot\br' \ ,
\ee
where $\dot\br,\dot\br'$ are the velocities in the corresponding Keplerian problem.
With classical computation, we obtain 
\be
\EQM{
\Frac{a}{r} \cos E = \Frac{2}{e} \sum_{k=1}^{+\infty} J_{k}(ke) \cos kM \cr
\Frac{a}{r} \sin E = 2 \sum_{k=1}^{+\infty} J'_{k}(ke) \sin kM 
}
\ee
where $E$ is the eccentric anomaly and $J_k(x)$ are the Bessel functions. The coordinates $(\dot X, \dot Y)$ of the velocity $\dot\br$ in the reference frame 
of the orbit with origin at perihelion are then easily expressed in Fourier series of the mean anomaly, as 
\be
\EQM{
\dot X = -na \left(\Frac{a}{r} \sin E\right) \cr
\dot Y =  na \sqrt{1-e^2} \left(\Frac{a}{r} \cos E\right) \ .
}
\ee
If we denote $\cZ = \dot X + \i \dot Y$, we then have 
for the spatial problem in the fixed reference frame, with the same notations as in the previous section\citep[][Eq. 37]{laskar_stability_1995},
\be
V= \Re\left(\tmu \cZ \bar\cZ' \e^{\i(\om-\om')} + \tnu\cZ\cZ' \e^{\i(\om+\om')}\right) \ .
\ee
If one considers the mutual inclination $J$, as in section \ref{sec_spa}, this 
expression simplifies since $\Om=\Om'$, $\si'=0$, $\co'=1$, 
and $\tmu, \tnu$ are then real, with  $\tmu=\cos^2(J/2)$, $\tnu= \sin^2(J/2)$. In  the planar case 
(Sect.  \ref{sec_plan}),  $\tmu=1$, $\tnu= 0$.

We  considered here heliocentric coordinates. One could  also use Jacobi coordinates for the three-body problem. 
In this case, 
the indirect part  does not require additional computations as it is expressed in term of $\uu\dpp\uu'$
\citep[see][]{Laskar:1990:a}. It should be 
noted that in terms of
both  heliocentric coordinates or  Jacobi coordinates, the indirect part does not contribute to the secular system.
We have provided here the expression of the indirect part for the computation of non-secular inequalities.

\section{Conclusion}
We have presented a self-contained exposition of the expansion of the three-body Hamiltonian in canonical  
heliocentric 
coordinates   in term of Legendre polynomials at any order in  the ratio of semi-major axes $\al$, 
which can also be adapted  in the case of Jacobi coordinates, where only the indirect part differs. 
We have included  here all the necessary material that allows one to write explicitly   the secular 
Hamiltonian at order $\al^{10}$ for the planar case, $\al^5$ for the spatial case expressed in terms
of the mutual inclination, and 
$\al^3$ for the spatial case in a fixed reference frame. With the additional computation of the required Hansen coefficients, 
the expressions of the Tisserand functions $F_n$ can also be used for a straightforward computation of the 
expression of a non-secular inequality $kM+k'M'$.

As the algorithms that are presented here are very simple, we have not added to this paper any
tables in electronic form.
Indeed, the reader who needs to use expressions of high order that do not appear in  the paper,  will 
have no problems in exploiting the algorithms given here  to derive the  required expressions. The 
written results of the paper can then be used to check his computations for the lowest orders. 
For example, the computation of $F_{100}$, in the spatial case  takes less than 8 minutes on an average 8 core desktop computer
in exact rational arithmetics using TRIP, but has 2343926 terms, while  $F_{50}$ needs only 7.5 seconds with 
 already  164151 terms. As the algorithms require  only a few lines of code (fewer than 20 in our case),  one understands that it is preferable to 
 compute the terms when they are needed than to store the values in electronic form.


\bibliography{../bib_secu}

\appendix

\onecolumn
\section{Computation of the Hansen coefficients}
\label{an_hans}
The Hansen coefficients $X_k^{n,m}(e)$ are defined as the Fourier coefficients 
\be
\left(\frac{r}{a}\right)^n \e^{\i mv}  = \sum_{k=-\infty}^{+\infty} X_k^{n,m}(e) \, \e^{\i kM} \ .
\ee
For arbitrary values of $k\in \mathbb{Z}$, the  Hansen coefficients  $X_k^{n,m}(e)$  can be expressed 
in an explicit manner as an infinite  series involving Bessel functions and 
hypergeometric functions \citep{Hansen_1855,hill_development_1875,Tisserand_1899}, or 
generalized Laplace coefficients
\citep{laskar_notegeneralized_2005}, but for $k = 0$, $X_k^{n,m}(e)$ reduces to  a  simple polynomial in $e$, $1/e$, $\sqrt{1-e^2}$, and $1/\sqrt{1-e^2}$. 
The literature on the computation of the Hansen coefficients has been  huge 
since the original work of \citet{Hansen_1855}, but we do not review it here.
We  concentrate  on the obtention of the coefficients $ X_0^{n,m}(e)$ and $ X_0^{-n,m}(e)$, 
for $n\geq 0$ and $0\leq m\leq n$ that are required to computate the averaged planetary or lunar Hamiltonian.
Using 
\be
dM = \Frac{r^2}{a^2\sqrt{1-e^2}} dv = \Frac{r}{a} dE
\ee
where $v$ is the true anomaly and $E$ the eccentric anomaly, we have for $n \geq 2$
\be
\EQM{
 X_0^{-n,m} &=
 \Frac{1}{\sqrt{1-e^2}} \Frac{1}{2\pi}\int_0^{2\pi} \left(\Frac{a}{r}\right)^{n-2} \e^{\i mv} dv \crm
 &=\Frac{1}{(1-e^2)^{n-3/2}} \Frac{1}{2\pi}\int_0^{2\pi} (1+e\cos v)^{n-2} \e^{\i mv} dv \crm
 &=\Frac{1}{(1-e^2)^{n-3/2}} \sum_{l=0}^{[(n-2-m)/2]}  \Frac{(n-2)!}{l!\,(m+l)!\,(n-2-(m+2l))!} \left(\Frac{e}{2}\right)^{m+2l} \ .
                }
\label{eq.anomv}
\ee
\tabd
We still need to consider the case $n=1$ for which the expansion in true anomaly $v$ is not suitable. We have immediately,  using (\ref{eq.anomE}) 
\be
X_0^{-1,0}=1 \ ; \qquad X_0^{-1,1} = \Frac{\sqrt{1-e^2}-1}{e} \ .
\ee
It is important to note that from (\ref{eq.anomv}), we have 
\be
X_0^{-n,m} = 0 \quad \hbox{for} \quad  n\geq 2 \quad \hbox{and} \quad  m \geq n-1 \ .
\ee
\tabc
The Hansen coefficients $X_0^{-n,m}$ are given in Table \ref{tab.hansn} for $n \geq 2$ and $m \leq n-2$.
On the other hand, the computation of $X_0^{n,m}(e)$ for $n \geq 0 $ is not as straightforward, as a direct expansion of 
\be
\EQM{
 X_0^{n,m} &=  \Frac{1}{2\pi}\int_0^{2\pi} \left(\Frac{r}{a}\right)^{n+1} \e^{\i mv} dE \crm
           &=  \Frac{1}{2\pi}\int_0^{2\pi} (1-e\cos E)^{n+1-m} (\cos E -e + i\sqrt{1-e^2}\sin E)^m dE \crm
\label{eq.anomE}
    }
\ee
in eccentric anomaly is far more complicated than the previous expression (\ref{eq.anomv}). 
One can still perform some formal expansion, 
but this would not be very useful, as it would contain expressions with many summations that cannot be easily 
reduced to a single summation as in (\ref{eq.anomv}). In fact, the only explicit computation  of 
$X_0^{n,m}(e)$ available in the literature, was obtained through a complex process  in \citep{Hansen_1855,hill_development_1875,Tisserand_1899},
using 
the auxiliary variable $\beta = ({1-\sqrt{1-e^2}})/{e}$. In this case, it can be shown that $X_0^{n,m}(e)$ is expressed as a finite 
hypergeometric series  in $\beta^2$. Using a relation among hypergeometric series due to Gauss, and the change of variable 
$e= 2\beta/(1+\beta^2)$, \citet{Tisserand_1899} provides  a finite expression of $X_0^{n,m}(e)$ in term of hypergeometric series 
of $e^2$. 
We  present here a more direct demonstration for the obtention of an explicit expression of $X_0^{n,m}(e)$ which is a recurrence 
using the relation among Hansen coefficients 
\be
X_0^{n,m}  = \Frac{1}{n+2-m} \left( \Frac{2}{e} (m-1) X_0^{n,m-1}  + (n+m) X_0^{n,m-2}  \right ) \ .
\label{eq.rec}
\ee
This relation appears in \citep{hughes_computation_1981}, but it was shown by \citet{laskar_notegeneralized_2005} that it 
is equivalent to a recurrence relation obtained on Laplace coefficients by \citet{Laplace_1785}.
The computation of $X_0^{n,0}$  for $n\geq 0$ using (\ref{eq.anomE}) is straightforward and gives 
\be
 X_0^{n,0} = \sum_{l=0}^{[(n+1)/2]}  \Frac{(n+1)!}{l!\,l!\,(n+1-2l)!} \left(\Frac{e}{2}\right)^{2l}\ .
\ee
The computation of $X_0^{n,1}$  can also be performed using (\ref{eq.anomE}). Changing $E$ to $-E$, it is immediate to see 
that the part in $\sin E $ cancel, and we are left with 
\be
 X_0^{n,1} =- (n+2)\sum_{l=0}^{[n/2]}  \Frac{n!}{l!\,(l+1)!\,(n-2l)!} \left(\Frac{e}{2}\right)^{2l+1} \ .
\ee
We can now prove by recurrence  using the relation (\ref{eq.rec}) that the general form of   $X_0^{n,m}$ for $n\geq 0$ and $0\leq m \leq n$ is  
\be
 X_0^{n,m} =(-1)^m \Frac{(n+ 1+m)!}{(n+1)!} \sum_{l=0}^{[(n+1-m)/2]}  \Frac{(n+1-m)!}{l!\,(m+l)!\,(n+1- m-2l)!} \left(\Frac{e}{2}\right)^{m+2l} \ .
 \label{eq.xnm}
\ee
The computation is delicate but straightforward. One can  first show that the two first elements (for $l=0$) of the polynomial expressions 
of $ X_0^{n,m-1}$ and $ X_0^{n,m-2}$ in  (\ref{eq.rec}) cancel. We then change the index from $l$ to $l+1$ and show that the 
general term of the sum in  (\ref{eq.rec})  gives the general term of $ X_0^{n,m}$, and  that the
last term  of $ X_0^{n,m-2}$ will give the last term  of  $ X_0^{n,m}$. 
The expression (\ref{eq.xnm}) is  equivalent to the expression of \citep{Tisserand_1899}. 
\nul{
\be
 X_0^{n,m} =(-\Frac{e}{2})^m \Frac{(n+m+1)!}{m!(n+1)!}F(\Frac{ m-n-1}{2},\Frac{m-n}{2},m+1,e^2) \ 
\ee
}
The expressions of the Hansen coefficients $X_0^{n,m}$ are given in
Table \ref{tab.hans} for $0\leq n\leq 10$ and $0\leq m\leq n$.

\pagebreak
\section{Computation of Tisserand polynomials and Tisserand functions}
\label{sec.func_tisserand}

Expansions valid for all inclinations were already given by \citet{Hansen_1855}, and \citet{tisserand_memoire_1885} in his  
researches on asteroidal motions,
but these derivations are rather complex. 
\mybf{Although in (\ref{sec.simpl}) we  present some of Tisserand's computations, 
we  make first, as in the planar case some direct expansions and  
reordering of terms  to obtain 
expressions that are as compact as possible.} 

\subsection{Tisserand functions with respect to mutual inclination}
\label{sec.mutual}

With $x=u-u'$, $y=u+u'$, $\mu= \cos^2 (J/2)$, $\nu= \sin^2 (J/2)$, we have
\be
\uu\dpp\uu' = \mu \cos x + \nu \cos y \ ,
\label{eq.uuxy}
\ee
and with a straightforward expansion in complex notation, we   obtain
\be
(\uu\dpp\uu')^m = \Frac{m!}{2^m} \sum_{ \underset{0\leq k_i\leq m}{ k_1+k_2+k_3+k_4=m}} \Frac{\mu^{k_1+k_2}\nu^{k_3+k_4}}{k_1!k_2!k_3!k_4!}
\e^{\i x(k_1-k_2)}\e^{\i y(k_3-k_4)} \ ,
\ee
which we can reorder as 
\be
(\uu\dpp\uu')^m = \Frac{m!}{2^m} \sum_{r=0}^{m} \sum_{p=-r}^{m-r} \cC_{p,m-2r-p}^{(m)}\, \e^{\i x\,p}\e^{\i y\,(m-2r-p)}
\ee
with
\be
\cC_{p,m-2r-p}^{(m)} = \sum_{l=\max(0,-p)}^{\min(r,m-r-p)} \Frac{\mu^{2l+p}\nu^{m-2l-p}}{l!(l+p)!(r-l)!(m-p-r-l)!}\ .
\label{eq.Cmpq}
\ee
With the expansion (\ref{eq.fn}), we have then
\be
F_n =  \sum_{k=0}^{[n/2]} q_{n,k}\,   
\sum_{r=0}^{n-2k} \sum_{p=-r}^{n-2k-r} \cC_{p,n-2k-2r-p}^{(n-2k)}\, \e^{\i x\,p}\e^{\i y\,(n-2k-2r-p)}
\label{eq.fkrp}
\ee
with 
\be
q_{n,k}=p_{n,k}\, \Frac{(n-2k)!}{2^{n-2k}} = \frac{(-1)^k}{2^{(2n-2k)}} \frac{(2n-2k)!}{k! (n-k)!} \ .
\label{eq.qnk}
\ee

We can then reorder (\ref{eq.fkrp}) with $s=k+r$ as 
\be
F_n =     
\sum_{s=0}^{n} \sum_{k=0}^{\min(s,n-s)}\sum_{p=k-s}^{n-k-s}  q_{n,k}\,  \cC_{p,n-2s-p}^{(n-2k)}\, \e^{\i x\,p}\e^{\i y\,(n-2s-p)} \ ,
\label{eq.fskp}
\ee
exchange the sum in $p$ and $k$
\be
F_n =    
\sum_{s=0}^{n} \sum_{p= -s}^{n -s}  \sum_{k=0}^{\min(s+p,n-s-p,s,n-s)} q_{n,k}\,  \cC_{p,n-2s-p}^{(n-2k)}\, \e^{\i x\,p}\e^{\i y\,(n-2s-p)}
\label{eq.fskpii}
\ee
and set $q=p+s$ which gives 
\be
F_n =    
\sum_{s=0}^{n} \sum_{q= 0}^{n }  \left( \sum_{k=0}^{\min(q,n-q,s,n-s)} q_{n,k}\,  \cC_{q-s,n-q-s}^{(n-2k)}\right) \, \e^{\i x\,(q-s)}\e^{\i y\,(n-q-s)} \ .
\label{eq.fskpiiii}
\ee
That is, with $x= u-u'$ and $y=u+u'$, 
\be
F_n =  
\sum_{s=0}^{n} \sum_{q= 0}^{n }  \cQ^{(n)}_{s,q} (\mu,\nu) \, \e^{\i u\,(n-2s)}\e^{\i u'\,(n-2q)}
\label{eq.fnuuB}
\ee
with
\be
\cQ^{(n)}_{s,q}(J) = \sum_{k=0}^{\min(q,n-q,s,n-s)} q_{n,k}\,  \cC_{q-s,n-q-s}^{(n-2k)} \ ,
\label{eq.tis}
\ee
that is 
\be
\cQ^{(n)}_{s,q}(J) = \sum_{k=0}^{\min(q,n-q,s,n-s)} q_{n,k}\, 
\sum_{l=\max(0,s-q)}^{\min(s-k,n-k-q)} \Frac{\mu^{2l+q-s}\nu^{n-2k-q+s-2l }}{l!(l+q-s)!(s-k-l)!(n-k-q-l)!} \ .
\label{eq.tisi}
\ee
 
\subsection{Tisserand functions with respect to a fixed reference frame}
\label{sec.reference}

\newcommand{\bmus}{\overline{\mu}_*}
\newcommand{\bnus}{\overline{\nu}_*}

In the case of a fixed reference frame (Sect.\ref{sec_spaii}), 
the expressions  differ slightly, but are more complicated. 
We have (\ref{eq.uuii})
\be
\uu\dpp\uu'  =  a\cos x -a'\sin x   + b\cos y -b'\sin y \ ,
\ee
that is, with $\tmu = a + i a'$, $\tnu = b+ib'$, 
\be
\uu\dpp\uu'  = \Frac{1}{2}\left[ \tmu \e^{\i x} + \bmus \e^{-\i x}+ \tnu \e^{\i y} + \bnus \e^{-\i y}\right] \ .
\ee
This is similar to the previous expression (\ref{eq.uuxy}), and the computation of the Tisserand functions $F_n$ will 
follow the same lines and gives the more general expression 

\be
F_n =  
\sum_{s=0}^{n} \sum_{q= 0}^{n }  \tilde{\cQ}^{(n)}_{s,q}  \, \e^{\i u\,(n-2s)}\e^{\i u'\,(n-2q)}
\label{eq.fnuuii}
\ee
with 
\be
\tilde{\cQ}^{(n)}_{s,q}(\tmu,\tnu)  = \sum_{k=0}^{\min(q,n-q,s,n-s)} q_{n,k}\, 
\sum_{l=\max(0,s-q)}^{\min(s-k,n-k-q)} \Frac{\tmu^{l+q-s}\,\bmus^{\ l}\,\tnu^{n-k-q-l }\,\bnus^{\ s-k-l }}{l!(l+q-s)!(s-k-l)!(n-k-q-l)!} \ .
\label{eq.tisii}
\ee
The coefficients $\cQ^{(n)}_{s,q}(J)$ (\ref{eq.tisi}) and
$\tilde{\cQ}^{(n)}_{s,q}(\tmu,\tnu)$ (\ref{eq.tisii}) in the Tisserand functions
are very similar.
Indeed, using $\mu=1-\nu$ and $\abs{\tmu}=1-\abs{\tnu}$ (\ref{eq.tnutmu}), we have
\be
\cQ^{(n)}_{s,q}(J) = \mu^{q-s}\nu^{n-q-s}A^{(n)}_{q-s,n-q-s}(\nu),
\qquad
\tilde{\cQ}^{(n)}_{s,q}(\tmu,\tnu) =
\tmu^{q-s}\tnu^{n-q-s}A^{(n)}_{q-s,n-q-s}(\abs{\tnu}),
\label{eq.Ansq}
\ee
with $\nu=\abs{\tnu}=\sin^2(J/2)$ and
\be
A^{(n)}_{q-s,n-q-s}(x) = \sum_{k=0}^{\min(q,n-q,s,n-s)} q_{n,k}\,
\sum_{l=\max(0,s-q)}^{\min(s-k,n-k-q)}
\Frac{(1-x)^{2l}x^{-2k+2s-2l}}{l!(l+q-s)!(s-k-l)!(n-k-q-l)!} \ .
\label{Anqs}
\ee

\subsection{Tisserand simplification }
\label{sec.simpl}

The above expression (\ref{Anqs})  is a double sum. 
In the mutual inclination case, 
\citet{tisserand_memoire_1885}
could reduce it to a single sum using hypergeometric functions. 
He first considers the differential equation
satisfied by Legendre polynomials
\be
(1-z^2)\frac{d^2P_n}{dz^2}-2z\frac{dP_n}{dz}+n(n+1)P_n = 0 \ .
\label{eq.diffPn}
\ee
As $F_n=P_n(z)$ (Eq.\ref{eq.fn}) with $z=\cos x + \nu (\cos y - \cos x)$   
 (Eq. \ref{eq.uuxy}), we have from (\ref{eq.diffPn}),
\be
\nu(1-\nu)\Dron{^2F_n}{\nu^2}+(1-2\nu)\Dron{F_n}{\nu}
+\frac{1}{1-\nu}\Dron{^2F_n}{x^2}+\frac{1}{\nu}\Dron{^2F_n}{y^2}
+n(n+1)F_n=0\ .
\label{eq.diffFn}
\ee
Using expressions (\ref{eq.fnuuB}, \ref{eq.Ansq}), one then replaces $F_n$  in (\ref{eq.diffFn}) by 
\be
F_n = \sum_k\sum_l(1-\nu)^k\nu^lA^{(n)}_{k,l}(\nu)\e^{\i kx}\e^{\i ly} \ ,
\ee
which leads to
\be
\sum_k\sum_l(1-\nu)^k\nu^l\cA^{(n)}_{k,l}(\nu)\e^{\i kx}\e^{\i ly}=0 \ ,
\label{eq.diffFnii}
\ee
where
\be
\cA^{(n)}_{k,l}(\nu) = \nu(1-\nu)\frac{d^2A^{(n)}_{k,l}}{d\nu^2}
+\big[1+2l-2(k+l+1)\nu\big]\frac{dA^{(n)}_{k,l}}{d\nu}
+(n-k-l)(n+k+l+1)A^{(n)}_{k,l}\ .
\label{eq.diffAn}
\ee
The equality (\ref{eq.diffFnii}) must be satisfied for all $x$ and $y$,
thus all the coefficients $\cA^{(n)}_{k,l}(\nu)$ (\ref{eq.diffAn}) are
equal to 0. The solutions of (\ref{eq.diffAn}) are
\be
A^{(n)}_{k,l}(\nu) = K^{(n)}_{k,l}F(k+l-n,k+l+n+1;1+2l;\nu) 
\label{eq.solAn}
\ee
or equivalently
\be
A^{(n)}_{q-s,n-q-s}(\nu) =
K^{(n)}_{q-s,n-q-s}F(-2s,2n-2s+1;2n-2q-2s+1;\nu) \ ,
\label{eq.solAnii}
\ee
where $K^{(n)}_{k,l}$ is an unknown constant and $F$ an hypergeometric
function \citep[e.g.][]{Whittaker:1927:a}. Let us assume that the quantity $2n-2q-2s+1$ is positive. If it
is not the case, one can make the change of variable
$(s,q)\rightarrow(n-s',n-q')$. Then $2n-2q'-2s'+1$ is positive and since
$F_n$ is real, $\cQ^{(n)}_{s,q}(\nu)=\cQ^{(n)}_{s',q'}(\nu)$
(\ref{eq.fnuuB}). From (\ref{eq.solAnii}), one thus has
\be
A^{(n)}_{q-s,n-q-s}(\nu) = K^{(n)}_{q-s,n-q-s}
\frac{(2s)!(2n-2q-2s)!}{(2n-2s)!}\sum_{k=0}^{2s}(-1)^k
\frac{(2n-2s+k)!}{(2s-k)!(2n-2q-2s+k)!}\frac{\nu^k}{k!}\ .
\label{eq.An}
\ee
\citet{tisserand_memoire_1885} needs then some lengthy computations 
to determine $K^{(n)}_{k,l}$ from the
coefficient of   $\nu^n$ in $F_n$.
Here, we use the expression (\ref{eq.tisi}) with (\ref{eq.Ansq}) and 
(\ref{eq.An}). With $\mu=1-\nu$, we get
\be
K^{(n)}_{q-s,n-q-s}(-1)^{q-s}\frac{(2n-2q-2s)!(2n)!}{(2n-2s)!(2n-2q)!}
=(-1)^{q-s}\frac{(2n)!}{2^{2n}n!}\sum_{l=\max(0,s-q)}^{\min(s,n-q)}
\frac{1}{l!(l+q-s)!(s-l)!(n-q-l)!}\ .
\label{eq.Kn}
\ee
Calculating the coefficient
of the term $x^s$ in $(1+x)^{n-q}(1+x)^q=(1+x)^n$, one finds
\be
\sum_{l=\max(0,s-q)}^{\min(s,n-q)}\frac{1}{l!(l+q-s)!(s-l)!(n-q-l)!}=
\frac{n!}{s!(n-s)!q!(n-q)!}\ .
\ee
Thus, from (\ref{eq.Kn}), we   obtain
\be
K^{(n)}_{q-s,n-q-s} =
\frac{1}{2^{2n}}\frac{(2n-2s)!(2n-2q)!}{s!(n-s)!q!(n-q)!(2n-2q-2s)!}\ .
\label{eq.Knqs}
\ee
Finally,  the most general case,  where inclinations are defined with
respect to a fixed reference plane, can be as well derived from  (\ref{eq.Ansq})
and (\ref{eq.Knqs}). We have
\be
F_n = \sum_{s=0}^n\sum_{q=0}^n\tilde{\cQ}^{(n)}_{s,q}(\tmu,\tnu)
\e^{\i(n-2s)u}\e^{\i(n-2q)u'} \ ,
\ee
where
\be
\tilde{\cQ}^{(n)}_{s,q}(\tmu,\tnu) = \left\{
\EQM{
\tmu^{q-s}\,\tnu^{n-q-s}A^{(n)}_{q-s,n-q-s}(\abs{\tnu})
&\quad \text{if\ } s+q\leq n\ , \crm
\bmus^{\ s-q}\,\bnus^{\ s+q-n}A^{(n)}_{s-q,s+q-n}(\abs{\tnu})
&\quad\text{else,}
}\right.
\label{eq.Qtilde}
\ee
and
\be
A^{(n)}_{q-s,n-q-s}(x) = \frac{1}{2^{2n}}
\frac{(2s)!(2n-2q)!}{s!(n-s)!q!(n-q)!}
\sum_{k=0}^{2s}(-1)^k\frac{(2n-2s+k)!}{(2s-k)!(2n-2q-2s+k)!}
\frac{x^k}{k!}\ .
\ee
In (\ref{eq.Qtilde}), we used the fact that $\tilde{\cQ}^{(n)}_{s,q}$ is the complex
conjugate of $\tilde{\cQ}^{(n)}_{n-s,n-q}$ since $F_n$ is real.

\end{document}